\documentclass[12pt]{article}
\usepackage{amsmath,amsfonts,amssymb}
\usepackage{epsfig}
\usepackage{array}
\def\sectappend#1
{\vskip35pt plus12pt
{\raggedright
\noindent {\Large \bf {#1}\par}}
\vskip13pt}

\def\build#1_#2^#3{\mathrel{\mathop{\kern 0pt#1}\limits_{#2}^{#3}}}

\def\beq{\begin{equation}}
\def\eeq{\end{equation}}
\def\beqn{\begin{eqnarray}}
\def\eeqn{\end{eqnarray}}

\expandafter\ifx\csname amssym.def\endcsname\relax \else\endinput\fi

\expandafter\edef\csname amssym.def\endcsname{
       \catcode`\noexpand\@=\the\catcode`\@\space}

\catcode`\@=11

\def\undefine#1{\let#1\undefined}
\def\newsymbol#1#2#3#4#5{\let\next@\relax
 \ifnum#2=\@ne\let\next@\msafam@\else
 \ifnum#2=\tw@\let\next@\msbfam@\fi\fi
 \mathchardef#1="#3\next@#4#5}
\def\mathhexbox@#1#2#3{\relax
 \ifmmode\mathpalette{}{\m@th\mathchar"#1#2#3}
 \else\leavevmode\hbox{$\m@th\mathchar"#1#2#3$}\fi}
\def\hexnumber@#1{\ifcase#1 0\or 1\or 2\or 3\or 4\or 5\or 6\or 7\or 8\or
 9\or A\or B\or C\or D\or E\or F\fi}

\font\tenmsa=msam10 scaled \magstep1
\font\sevenmsa=msam7 scaled \magstep1
\font\fivemsa=msam5 scaled \magstep1
\newfam\msafam
\textfont\msafam=\tenmsa
\scriptfont\msafam=\sevenmsa
\scriptscriptfont\msafam=\fivemsa
\edef\msafam@{\hexnumber@\msafam}

\mathchardef\dabar@"0\msafam@39
\def\dashrightarrow{\mathrel{\dabar@\dabar@\mathchar"0\msafam@4B}}
\def\dashleftarrow{\mathrel{\mathchar"0\msafam@4C\dabar@\dabar@}}

\def\ulcorner{\delimiter"4\msafam@70\msafam@70 }
\def\urcorner{\delimiter"5\msafam@71\msafam@71 }
\def\llcorner{\delimiter"4\msafam@78\msafam@78 }
\def\lrcorner{\delimiter"5\msafam@79\msafam@79 }
\def\yen{{\mathhexbox@\msafam@55 }}
\def\checkmark{{\mathhexbox@\msafam@58 }}
\def\circledR{{\mathhexbox@\msafam@72 }}
\def\maltese{{\mathhexbox@\msafam@7A }}

\font\tenmsb=msbm10 scaled \magstep1
\font\sevenmsb=msbm7 scaled \magstep1
\font\fivemsb=msbm5 scaled \magstep1
\newfam\msbfam
\textfont\msbfam=\tenmsb
\scriptfont\msbfam=\sevenmsb
\scriptscriptfont\msbfam=\fivemsb
\edef\msbfam@{\hexnumber@\msbfam}

\def\widehat#1{\setbox\z@\hbox{$\m@th#1$}
 \ifdim\wd\z@>\tw@ em\mathaccent"0\msbfam@5B{#1}
 \else\mathaccent"0362{#1}\fi}
\def\widetilde#1{\setbox\z@\hbox{$\m@th#1$}
 \ifdim\wd\z@>\tw@ em\mathaccent"0\msbfam@5D{#1}
 \else\mathaccent"0365{#1}\fi}

\font\teneufm=eufm10 scaled \magstep1
\font\seveneufm=eufm7 scaled \magstep1
\font\fiveeufm=eufm5 scaled \magstep1
\newfam\eufmfam
\textfont\eufmfam=\teneufm
\scriptfont\eufmfam=\seveneufm
\scriptscriptfont\eufmfam=\fiveeufm

\csname amssym.def\endcsname

\def\ben{\begin{enumerate}}
\def\een{\end{enumerate}}
\def\bit{\begin{itemize}}
\def\eit{\end{itemize}}

\def\b{\beta}
\def\g{\gamma}

\def\Reel{\mathop{\rm I\! R}}

\date{} % this removes the date of the day !!!

\def\lsim{\raise0.3ex\hbox{$<$\kern-0.75em\raise-1.1ex\hbox{$\sim$}}}
\def\gsim{\raise0.3ex\hbox{$>$\kern-0.75em\raise-1.1ex\hbox{$\sim$}}}

\title{\large \bf
Cluster Percolation in\\
O(n) Spin Models}

\author{\normalsize
Ph. Blanchard$^{1}$,
S. Digal$^{1}$,
S. Fortunato$^{1}$,\\
\normalsize D. Gandolfo$^{2}$, 
T. Mendes$^{1}$,
H. Satz$^{1}$
}

\begin{document}
\maketitle
\begin{quote}
\footnotesize
\begin{center}
$^{1}$ Fakult\"{a}t f\"{u}r Physik, 
Universit\"{a}t Bielefeld, D-33615, Bielefeld, Germany.\\
$^{2}$ D\'ept. de Math., Universit\'e de 
Toulon et du Var, , F-83957 La Garde Cedex,\\ 
\& CPT, CNRS, Luminy, case 907, 13288 Marseille Cedex 09, FRANCE.
\end{center}
\end{quote}

\vskip 1.0cm

\noindent

\centerline{\bf Abstract:}

\medskip

The spontaneous symmetry breaking in the Ising model
can be equivalently described in terms of
percolation of Wolff clusters. 
In O(n) spin models similar clusters can be
built in a general way, and they are currently used
to update these systems in Monte Carlo 
simulations. We show that 
for 3-dimensional O(2), O(3) and O(4) such clusters are indeed 
the physical `islands' of the systems, i.e., they percolate
at the physical threshold and the percolation exponents
are in the universality class of the corresponding model.  
For O(2) and O(3) the result is proven analytically, for 
O(4) we derived it by numerical simulations.
  
\vskip 1cm      

\section{Introduction}

The possibility to interpret the critical 
behaviour of dynamical systems in terms of
percolation of geometrical structures 
of the system has always had a great appeal 
in the study of critical phenomena \cite{fisher} - \cite{CK}.
Attempts in this direction were already done in the 70's,
when one began to study the behaviour of 
clusters of nearest-neighbour like-signed spins in the Ising model.
It turned out that in two dimensions these elementary
site percolation clusters indeed undergo a geometrical 
transition exactly at the critical threshold of 
the Ising model \cite{Coni}. This result, which is not valid in three
dimensions \cite{Krum}, is anyway not so appealing, because 
the critical exponents derived by the percolation 
variables do not coincide with the Ising ones \cite{Sykes}. The
correspondence
between the geometrical and the thermal 
phenomenon is therefore only partial.

The problem was solved by A. Coniglio and W. Klein
\cite{CK}, making use of a different
definition for the clusters. Such definition had already
been used by Fortuin and Kasteleyn to show that
the partition function of the Ising 
model can be rewritten in purely geometrical terms
as a sum over clusters configurations \cite{FK}.
According to the Fortuin-Kasteleyn prescription,
two nearest-neighbouring spins of the same
sign belong
to the same cluster with a probability $p=1-exp(-2{\beta})$
($\beta=J/kT$, J is the Ising coupling). Coniglio and Klein
\cite{CK} showed that the
geometrical transition of
these clusters leads to the required 
critical indices of the Ising model, threshold and exponents.

The clusters  
of the Monte Carlo cluster update introduced by U. Wolff \cite{Wo}
for O(n) spin models coincide 
with the Fortuin-Kasteleyn 
ones when n=1 (which is just the Ising model).
The O(n) models 
without external field in three space dimensions ($n{\geq}2$)
undergo a phase transition
due to the spontaneous breaking of the continuous rotational
symmetry of their Hamiltonian.
Such models are very interesting: some physical systems
in condensed matter physics are directly
associated to them. 
The three-dimensional O(3) model is the low-temperature 
effective model for a bidimensional quantum antiferromagnet
\cite{mano}. The O(2) model in three dimensions is known to be
in the same universality class as superfluid ${^4}He$.
%But the 
O(n) models are also very useful to 
study relativistic field theories.
The O(4) model in three dimensions has
been conjectured to be in the same universality class 
as the finite-temperature chiral phase transition of QCD with
two flavours massless quarks \cite{wilczeck}.

The general definition of Wolff clusters for O(n) spin models
inspired this work. Can one describe the critical behaviour
of O(n) without field 
in three dimensions in terms of the percolation of these
clusters, like in the case n=1?
We will show that this is indeed true at least for
O(2), O(3) and O(4). 
The fact that the Wolff 
clusters percolate at the physical critical point 
was recently proven analytically for O(2) and O(3) \cite{Cha1, Cha2}.
Although nothing about the exponents was mentioned, 
we will show that starting from some relations
established in \cite{Cha1, Cha2} 
it is also possible to deduce the equality of the 
critical exponents for O(2) and O(3). 
%Before finding that we had already began 
%to perform computer simulations on O(2) and O(4)
%in order to prove the result numerically, and 
%we decided to carry them out anyway up to the end.
%Therefore we will present a numerical analysis  
%also for O(2), although it wouldn't be necessary
%to show the validity of the result.
%For O(4) instead, we can only rely on the results
%of the simulations.
We have also performed computer simulations on O(2) and O(4)
in order to illustrate this result for O(2), and to prove it
numerically for O(4).
  
\section{O(n) models and Wolff clusters}

The O(n) spin models with no 
external magnetic field have the following Hamiltonian:

\begin{equation}
H=-J\sum_{\langle{{\bf i},{\bf j}}\rangle}{\bf s_i s_j},
\end{equation}

\noindent where $i$ and $j$ 
are nearest-neighbour sites on a d-dimensional 
hypercubic lattice, and 
${\bf s_i}$ is an n-component unit vector at site $i$ ($J$ is the 
coupling). The partition function
of these models at the temperature $T$ is 
\begin{equation}
Z(T) = \int{\cal D}[{\bf s}] \exp\{ \beta \sum_{\langle {{\bf i},{\bf j}} \rangle}{\bf s_i s_j} \} \label{1}
\end{equation}
where $\beta=J/kT$ and the integral is extended over all spin configurations
$\{{\bf s}\}$ of the system.
In three dimensions the O(n) models undergo a second-order 
phase transition. The order parameter of this transition
is the normalized magnetization $M={{\frac{1}{V}}\sum_{\bf i}{\bf s_i}}$
(V is the lattice volume). 

Numerical simulations of O(n) models became much 
quicker and more 
effective after U. Wolff \cite{Wo} introduced a Monte Carlo
algorithm based on simultaneous updates of large clusters of spins, 
generalizing the Swendsen Wang algorithm \cite{SW}
to the continuous-spin case.
This algorithm has the remarkable advantage that it eliminates
the problem of critical slowing down, 
an effect that makes simulations around criticality
very lengthy with traditional local methods
(Metropolis, heat bath).
The 
Wolff algorithm can be basically divided in two 
phases:

1) a cluster of spins is 
% built up according to some rules;
selected;

2) the spins of this cluster are "flipped", i.e. they 
are reflected
with respect to some defined hyperplane.

For details of the flipping procedure, see \cite{Wo}.
Here we are interested in the way to build up the clusters.
We can split this procedure in two steps:

a) choose a random n-component unit vector {\bf r};
%vector {\bf r} of O(n);

b) bind
together pairs of nearest-neighbouring sites i,j with
the probability 
\begin{equation}
p(i,j)=1-exp\{min[0,-2{\beta}({\bf s_i}{\cdot}{\bf r})
({\bf s_j}{\cdot}{\bf r})]\}.
\end{equation}
From this prescription it follows that if 
the two spins at two nearest-neighbouring 
sites $i$ and $j$ are such that their projections onto the random vector 
{\bf r} are of opposite signs, they will never belong
to the same cluster ($p(i,j)=0$). The random vector {\bf r}, therefore, divides
the spin space in two hemispheres, separating the spins
which have a positive projection onto it from the ones which have a negative
projection. The Wolff clusters are made out of spins which all lie
either in the one or in the other hemisphere.
In this respect, we can again speak of 'up' and 'down' spins, like
for the Ising model. In addition to that, the bond probability is
local, since it depends explicitly on the spin vectors 
${\bf s_i}$ and ${\bf s_j}$, and not only on the temperature
like the Fortuin-Kasteleyn factor.

The analogies with the Ising
model are however clear,
motivating the attempt to study the percolation properties of these
clusters.
% and the attempt to study
% the percolation properties of these clusters fully justified.

\section{Percolation exponents for O(2) and O(3)}

In \cite{Cha1,Cha2,SanGand} the random cluster representations of $O(n)$
models, $n=2,3$ have been derived (and exploited) through the Fortuin-Kasteleyn
transformation \cite{FK} of the Hamiltonians and similar results were
obtained in \cite{BCG} for the continuous (or classical) spin model \cite{Gr}.
Wolff random cluster probability distributions \cite{Wo} for these models
have been studied
and several monotonicity properties of these distributions (FKG properties)
\cite{FKG} have been established leading to the proof of the equivalence
between
the onset of magnetic ordering in the $O(n)$ models, $n=2,3$ and percolation
in the corresponding random Wolff cluster models.
From the results stated above follows in a natural way the
equality of the critical thermal and geometrical exponents.

In what follows we focus our attention on two
variables:
\begin{itemize}
%\noindent       {$\bullet$}  
\item The {\it percolation strength } $P$, 
      defined as the probability that a lattice site picked 
      up at random belongs to the percolating cluster. 
      $P$ is the {\it order parameter} of the percolation transition.

\item The {\it average cluster
        size } $S$, defined as 
      \setcounter{equation}{0}
      \begin{equation}
        S=\frac{\sum_{s} {{n_{s}s^2}}}{\sum_{s}{n_{s}s}}~.
        \label{S}
      \end{equation}
      Here, $n_{s}$ is the number of clusters of size $s$ per lattice site
      (i.e. divided by the lattice volume)
      and the sums exclude the percolating cluster (see \cite{stauff}).
\end{itemize}

We stress that in some cases one speaks of average cluster size referring
to the average size of {\it all} clusters, including the percolating
one. This definition makes the variable infinite above 
the percolation threshold. Below the threshold the two definitions
obviously coincide and therefore they share the same critical exponent. 

In the references \cite{Cha1,Cha2,SanGand,BCG}, it was proved that,
if $P(T)$ is the percolation
strength in the random Wolff cluster representation and $m(T)$ is the
magnetization in
the $O(n)$ spin system, then there exists a function $c(T)\in
C^{\infty}(\Reel^{+})$ such that
\setcounter{equation}{0}
\begin{equation}
P(T) \ge m(T) \ge c(T) \; P(T)
\label{pm}
\end{equation}
Similarly, denoting $S(T)$ the average size of the Wolff clusters
and $\chi(T)$ the (linear response) susceptibility, it was also
proved  that
\begin{equation}
S(T) \ge \chi(T) \ge c(T)^{2} S(T) 
\label{cs}
\end{equation}
where $c(T)$ is the same $C^{\infty}(\Reel)$ function as in (\ref{pm}).

From scaling theory (see \cite{B}), near criticality, the susceptibility is
believed to behave
according to the following law
\begin{equation}
\chi(T) \mathop{\sim}_{T\to T_c^{+}} (T-T_c)^{-\g}
\label{g1}
\end{equation}
and the average mean cluster size of Wolff clusters should follow the law
\begin{equation}
 S(T) \mathop{\sim}_{T\to T_c^{+}} \mid p(T)-p(T_c)\mid^{-\g'}
\label{g2}
\end{equation}
where $p(T)$ is the bond occupation probability in the Wolff random cluster
model (i.e. the Coniglio-Klein \cite{CK}  bond probability) given by $p(T)\sim
1-exp(-a/T)$, where $a$ does not depend on $T$.

From  \cite{Cha1,Cha2,SanGand,BCG} the equality of the critical
exponents $\g$ and
$\g'$ follows readily . 
Indeed, because of monotonicity, taking the logarithm in (\ref{g1}) and
(\ref{g2}), and using (\ref{cs}) one gets
\begin{equation}
-\g' \log \mid p(T)-p(T_c) \mid \; \ge -\g \log (T-T_c) \ge 2 \log c(T)
-\g' \log \mid
p(T)-p(T_c) \mid
\label{cs1}
\end{equation}
which reduces to
\begin{equation}
\g' \ge \g \; \frac{\log (T-T_c)}{\log \mid p(T)-p(T_c) \mid} \ge \g' -
\frac{2 \log c(T)}{\log \mid p(T)-p(T_c) \mid}
\label{cs2}
\end{equation}
Now when $T\to T_c^{+}$ and since $c(T)\in C^{\infty}(\Reel)$, the last term
vanishes and it is easy to see that
$\displaystyle \log (T-T_c) / \log \mid p(T)-p(T_c) \mid \mathop{\to}_{T\to
  T_c^{+}} 1$. When $T\to T_c^{+}$ we get 
the result $\g=\g'$.

Using (\ref{pm}) and the scaling
behaviours of the magnetization and the percolation strength in terms of
their
critical exponents $\b$ and $\b'$ respectively, one can show (following the
same
lines as before) that $\b=\b'$ when $T\to T_c^{+} \equiv p(T)\to
p(T_c)^{+}$, where  $p(T)$ is again the Coniglio-Klein bond probability
already  defined above.

Namely, the percolation strength is believed to behave \cite{B} as a function
of the elementary bond occupation probability $p$ according to the
following law
\begin{equation}
P(p)  \mathop{\sim}_{p\to p_c^{+}} (p-p_c)^{\b'} \qquad
p\equiv p(T)
\label{b1}
\end{equation}
whereas, the magnetization should behave as
\begin{equation}
m(T)  \mathop{\sim}_{T\to T_c^{-}} (T_c -T)^{\b}
\label{b2}
\end{equation}
then using the same procedure as below, we are led to the following expression
\begin{equation}
\b' \le \b \; \frac{\log (T_c-T)}{\log (p(T)-p(T_c))} \le \b' - \frac{\log
c(T)}{\log (p(T)-p(T_c))}
\label{mag1}
\end{equation}
which, using $\displaystyle \log (T_c - T) / \log (p(T)-p(T_c))
\mathop{\to}_{T\to
  T_c^{-}} 1$ and $\displaystyle \log c(T) / log (p(T)-p(T_c))
\mathop{\to}_{T\to
  T_c^{-}} 0$, gives
$$
\b=\b'.
$$
as claimed.

About eventual extensions of analytical results to other spin models,
we mention that the equivalence of the thermal and 
the geometrical phase transitions
is true for $Z(3)$ and $Z(4)$ models as a consequence of the equivalences
already established for the continuous spin Ising model and $O(2)$ \cite{BCG,Cha1}.
As for $O(n)$ models with $n>3$, we remark
that the proof for $O(3)$ was derived 
starting from the result for $O(2)$. However, 
some additional
conditions remain to be proved 
in order to be able to extend the result from $O(n-1)$ to $O(n)$ 
if $n>3$.

\section{Numerical Results}

We have investigated numerically the 3-dimensional
O(2) and O(4) models performing computer simulations for
several lattice sizes. 
The Monte Carlo update was performed by the Wolff algorithm, described
in Section 2. 
At the end of an iteration, 
the percolation strength $P$ and the average cluster size $S$ were
measured. This has been done for each of the models using
two different approaches.

\vskip 3mm
The {\bf first approach} is the traditional one, based on 
a complete analysis of the lattice configuration.
Once we have the configuration we want to analyze, we 
build Wolff clusters until all spins are set into clusters.
We assign to $P$ the value zero if there is no percolating cluster,
the ratio between the size of the percolating cluster
and the lattice volume otherwise.
We calculate $S$ using the standard formula (\ref{S}).
The operative definition of percolating cluster was
taken as follows. We say that a cluster percolates if it spans
the lattice from a face to the opposite one in each of
the three directions $x$, $y$, $z$. We made this choice to reduce
the possibility that, due to the finite size of the lattices,
one could find more than a spanning cluster making 
ambiguous the definition of our variables\footnote{In three dimensions 
  even this definition of spanning cluster
  does not exclude the possibility of having more than one
  of such clusters for the same configuration. 
  Nevertheless the occurrence of such cases 
  is so small that we can safely ignore them.}.  

In this approach we have used free boundary conditions.

\vskip 3mm
      The {\bf second approach} is based on a single-cluster
      analysis. Basically one studies the percolation
      properties of the cluster built during the update
      procedure. For the cluster building we 
      have considered periodic boundary conditions.
      Suppose that $s_c$ is the size of the cluster we built. 
      If it percolates,
      we assign value one to the strength $P$
      and zero to the size $S$; otherwise, we write
      zero for $P$ and $s_c$ for $S$.
      These definitions of $P$ and $S$ look different from 
      the standard definitions we have introduced above, but it is easy to see
      that they are instead equivalent to them.
      
      In fact, we build the cluster starting from a lattice site taken at
      random. In this way, the probability that 
      the cluster percolates (expressed by the new $P$)
      coincides with the probability that a site taken at random belongs
      to the percolating cluster (standard definition of $P$).
      As far as the average cluster size is concerned, we can repeat the same
      reasoning: the probability that the cluster we built is
      a non-percolating cluster of size $s_c$ is just 
      the probability $w_{s_c}$ that a randomly taken lattice site
      belongs to a non-percolating cluster of size $s_c$; $w_{s_c}$ is given by 
      \setcounter{equation}{0}
      \begin{equation}
        w_{s_c}\,=\,n_{s_c}\,{s_c}\,\,.
        \label{probw}
      \end{equation}
      Because of that, whenever we get a non-vanishing size 
      $s_c$, such value will be weighted by the probability 
      $w_{s_c}$ in the final average $S$, which is then given by the following
      formula:
      \begin{equation}
        S\,=\,\sum_{s_c}w_{s_c}s_c\,=\,\sum_{s_c}n_{s_c}\,{s_c}^2,
        \label{tereS}
      \end{equation}
      where the sum runs over the non-percolating clusters.
      We notice that Eq. (\ref{tereS}) coincides with 
      Eq. (\ref{S}), apart from the denominator $\sum_{s}n_{s}\,s$,
      which is just the density of the sites belonging to finite clusters.
      Since this term does not contribute to the divergence
      of the average cluster size, the power law behaviour
      of the two $S$'s at criticality is identical, 
      so that the critical exponent $\gamma$ is the same in both cases.
      
      As we have said, in the second approach we 
      select a single cluster at a time from the whole configuration. Because of that
      we have now some freedom of choosing the 
      definition of percolating cluster, as we do not risk, like in the first
      approach, to find more spanning structures.
      We say that the cluster percolates if
      it connects at least one face with the opposite one.

      In this way, also the definitions of percolating 
      clusters are different in the two approaches. This 
      certainly influences the results on finite lattices, but
      has no effects on the infinite-volume properties we are interested in.
      In fact, it is known that  one can have 
      at most a unique spanning cluster above the critical 
      density $p_c$ (in our case below the critical temperature $T_c$).
      Exactly at $p_c$ ($T_c$) there is a finite
      probability to have more than a spanning cluster
      \cite{aiz}. So, the two different
      definitions of percolating cluster we have
      adopted can lead to differences between the 
      infinite-volume values {\it only}
      at the critical point $p_c$ ($T_c$). But the critical exponents are, of course,
      not influenced by that, as they are determined by the behaviour
      of the percolation variables {\it near}
      the critical point, not exactly at $p_c$ ($T_c$).

      The second approach has the advantage that it does not
      require a procedure to reduce the configuration 
      of the system to a set of clusters; on the other hand,
      since it gets the information out of a single
      cluster, it requires a higher number of samples
      in order to measure the percolation
      variables with the same accuracy of the first method.
      Nevertheless, the iterations are faster due to the simpler
      measurement of observables, and are less correlated than in the
      first approach, since only a (random) limited region of the lattice  
      is considered in each sample.
%      It is not yet so clear which of the two methods is 
%      the more efficient, but for us it was more important
%      to compare results got in different ways.
      We find that both methods are efficient, and that it is important 
      to be able to compare results obtained in two such different ways.

For our numerical investigations 
we have also made use of another variable which can be extracted from the 
percolation strength $P$. On a finite lattice there is at any temperature
$\beta$ a well defined probability of having a spanning cluster.
We call it percolation cumulant and indicate it $\gamma_r$.
When the size of the lattice goes to infinity, $\gamma_r$ 
as a function of $\beta$ approaches a step function:
it is always zero below ${\beta}_{c}$ and always one above it.
To get the finite-size curves out of our measurements we must
basically see how often we found a percolating cluster ($P{\neq}0$) for a 
definite lattice size and a temperature $\beta$.

To evaluate the thresholds and the exponents
we have adopted finite-size-scaling techniques. 
We consider the general finite-size-scaling prediction for an observable
${\cal O}$
\begin{equation}
{\cal O}(t,L) \;=\; L^{\rho/\nu}\,Q_{\cal O}(L^{1/\nu}t)\;,
\label{scale}
\end{equation}
where $\,t=T-T_c\,$, $L$ is the 
linear dimension of the lattice, $\,Q_{\cal O}\,$ is a universal function and the
exponent $\rho$ is related to the critical behavior of ${\cal O}$
at infinite volume. Following the definitions given in Section 3,
we have $\,\rho = \gamma'\,$ for the observable $S$ 
and $\,\rho = -\beta'\,$ for $P$. 
%By plotting ${\cal O}$ as a function of $L$ at the critical temperature,
%which corresponds to $t = 0$, we can therefore obtain the value
%$\rho/\nu$
%directly from the slope in the graph.
For the percolation cumulant $\gamma_r$ we have $\rho=0$ \cite{B}, which
means that $\gamma_r$ curves
corresponding to different lattice sizes cross at the critical 
point: for $t=0$ and $\rho=0$, in fact, the observable 
of Eq. (\ref{scale}) is not $L$-dependent.    

Figs. 1 and 2 show $\gamma_r$ curves for 
O(2) and O(4), respectively. The agreement with
the physical thresholds (dashed lines) is clear.
Moreover, we could
already get indications about the class of critical exponents of
our clusters.
In fact, if one knows the critical point and the exponent $\nu$,
a rescaling of $\gamma_r$ as a function
of $(T-T_{c})L^{1/\nu}$ 
should give us the same function
for each lattice size (see Eq. \ref{scale}) . 
Figs. 3 and 4 show the rescaled percolation cumulant curves
for O(2), using $\beta_{c}=0.45416$
and two different values of the exponent $\nu$, respectively the
O(2) value and the random percolation one.

\vskip1cm

  \begin{picture}(95,150)
    \put(-15,3){\epsfig{file=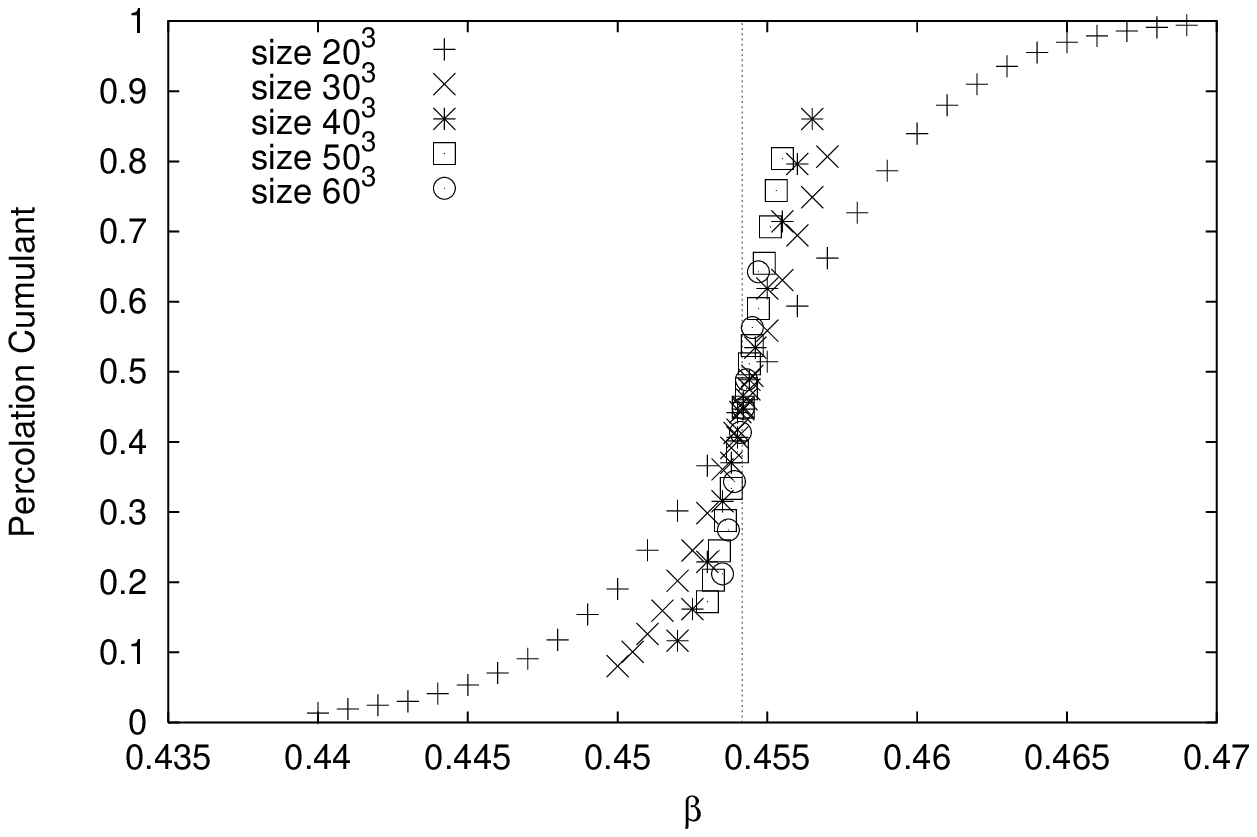,width=7.4cm}}
    \put(-15,-10){\begin{minipage}[t]{7.4cm}{{\footnotesize
          Figure 1. Percolation cumulant as function of $\beta$ for
          O(2) and five lattice sizes. The dashed line indicates the position
          of the thermal threshold \cite{O2}.}}
    \end{minipage}}
    \put(225,3){\epsfig{file=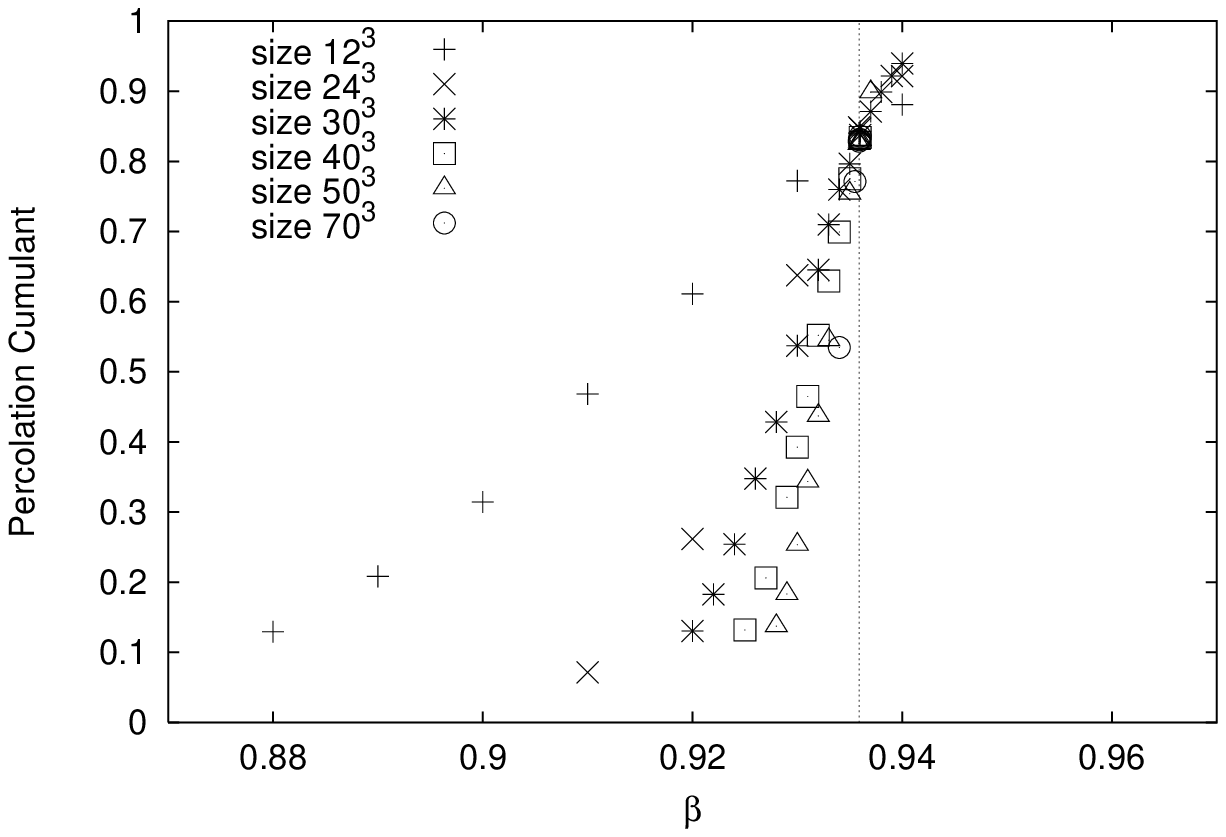,width=7.4cm}}
    \put(225,-10){\begin{minipage}[t]{7.4cm}{{\footnotesize
          Figure 2. Percolation cumulant as function of $\beta$ for
          O(4) and six lattice sizes. The dashed line indicates the position
          of the thermal threshold \cite{Manf}. }}
    \end{minipage}}
  \end{picture}

\vskip3cm

%Moreover, we could
%already get indications about the class of critical exponents of
%our clusters.
%In fact, if one knows the critical point and the exponent $\nu$,
%a rescaling of $\gamma_r$ as a function
%of $(T-T_{c})L^{1/\nu}$ 
%should give us the same function
%for each lattice size (see Eq. \ref{scale}) . 
%Figs.\ 3 and 4 show the rescaled percolation cumulant curves
%for O(2), using $\beta_{c}=0.45416$
%and two different values of the exponent $\nu$, respectively the
%O(2) value and the random percolation one.
The scaling we get in correspondence of the O(2) value is 
remarkable. In Figs. 5 and 6 we repeat the same
analysis for O(4) (${\beta}_{c}=0.9359$); also here 
it is clear that the percolation exponent $\nu$ is in agreement
with the O(4) value. (We have considered values for the O(2) and O(4)
models from Refs.\ \cite{O2} and \cite{Manf,O4} respectively.)

To determine more precisely the critical point we 
have used the scaling relation (\ref{scale}) for the variables $S$ and $P$.
By plotting ${\cal O}$ as a function of $L$ at the critical temperature,
we can obtain the exponents' ratio
$\rho/\nu$ directly from the slope of the 
data points in a log-log plot.

We concentrated 
ourselves on the critical regions that we localized through the 
percolation cumulant and performed more simulations for several 
$\beta$
values looking for the $\beta$'s
for which we get the best $\chi^2$
for the linear fit of the data points in a log-log plot.
The results for 
O(2) and O(4) are in Tables
I and II, respectively. 
  
\vskip1cm
\begin{picture}(95,160)
    \put(-15,3){\epsfig{file=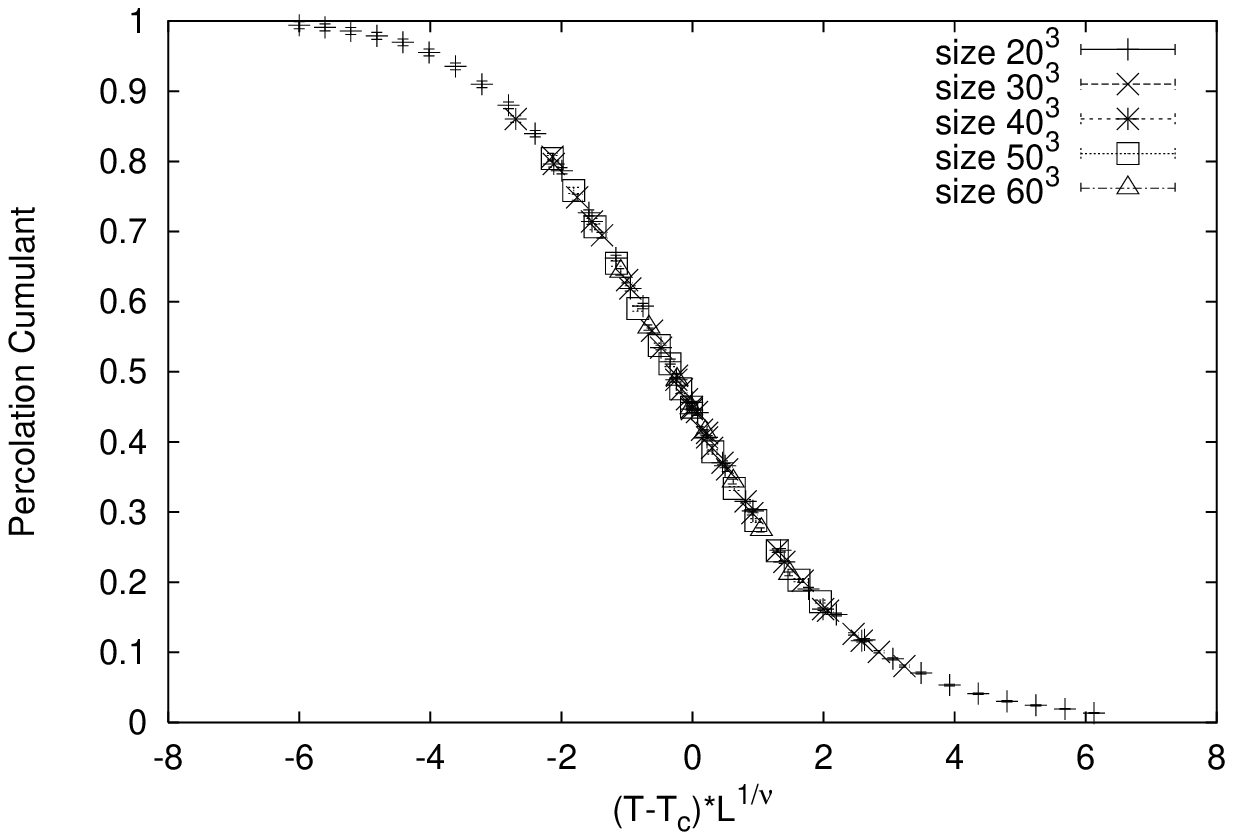,width=8.5cm}}
    \put(-15,-10){\begin{minipage}[t]{7.5cm}{{\footnotesize
          Figure 3. Rescaled percolation cumulant for O(2)
          using $\beta_{c}=0.45416$ and the
          O(2) exponent $\nu=0.672$.}}
    \end{minipage}}
    \put(220,3){\epsfig{file=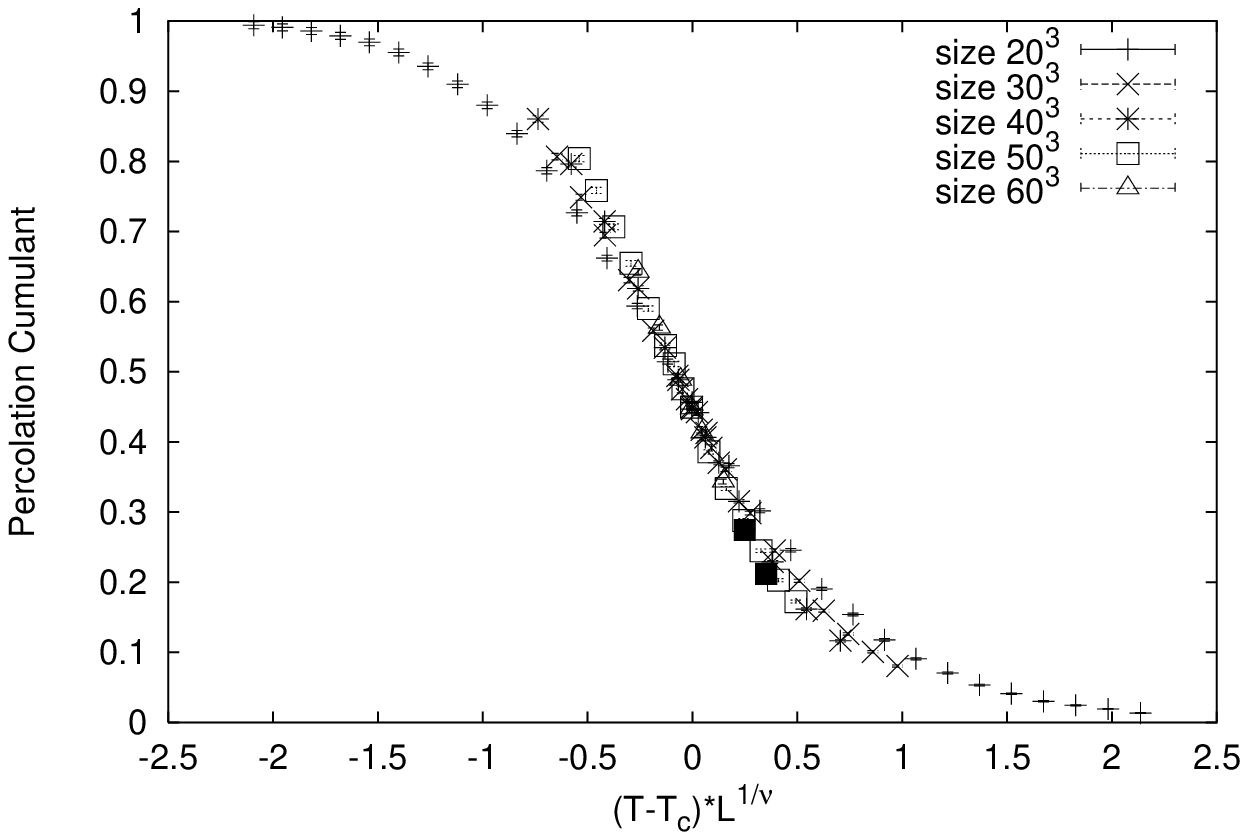,width=8.5cm}}
    \put(220,-10){\begin{minipage}[t]{7.5cm}{{\footnotesize
          Figure 4. Rescaled percolation cumulant for O(2)
          using $\beta_{c}=0.45416$ and the
          3-dimensional random percolation exponent $\nu=0.88$.}}
    \end{minipage}}
  \end{picture}
\vskip2cm
  \begin{picture}(95,180)
     \put(-15,3){\epsfig{file=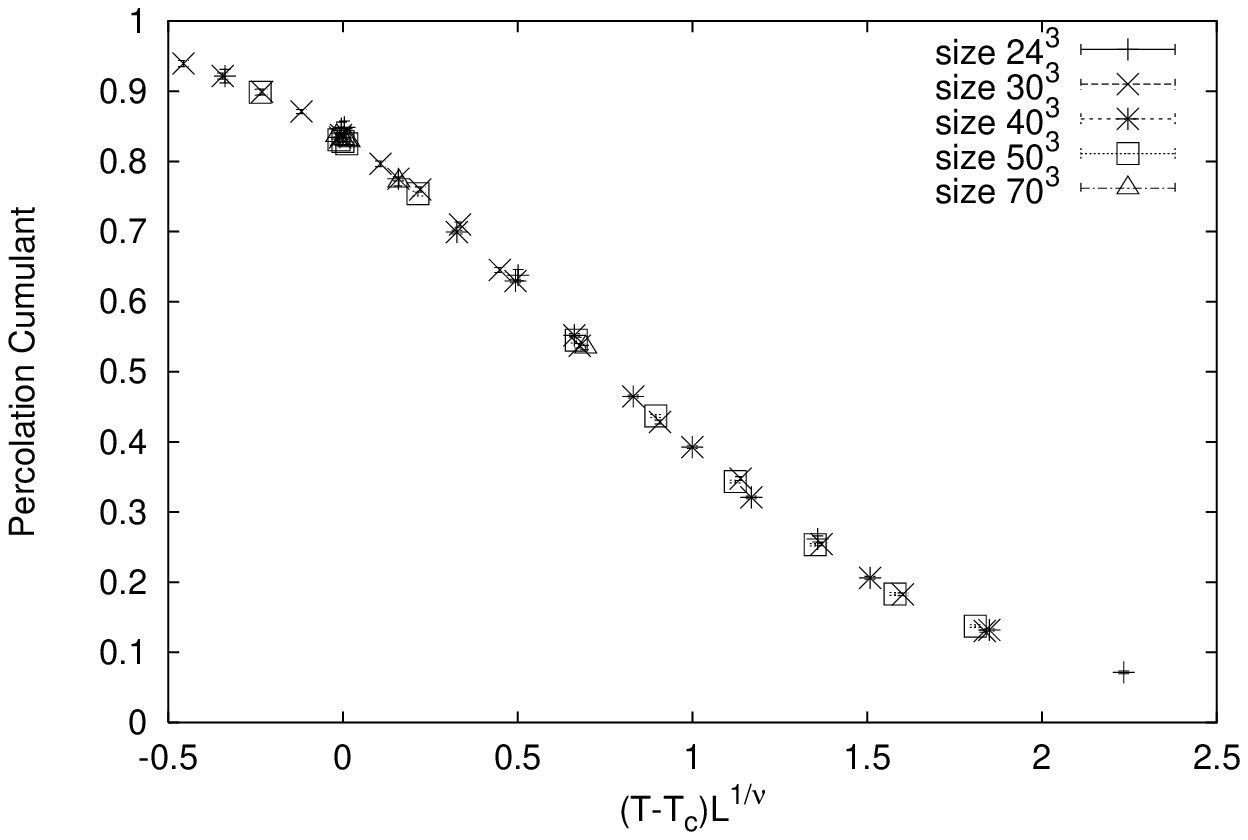,width=8.5cm}}
    \put(-15,-10){\begin{minipage}[t]{7.5cm}{{\footnotesize
          Figure 5. Rescaled percolation cumulant for O(4)
          using $\beta_{c}=0.9359$ and the
          O(4) exponent $\nu=0.742$.}}
    \end{minipage}}
    \put(220,3){\epsfig{file=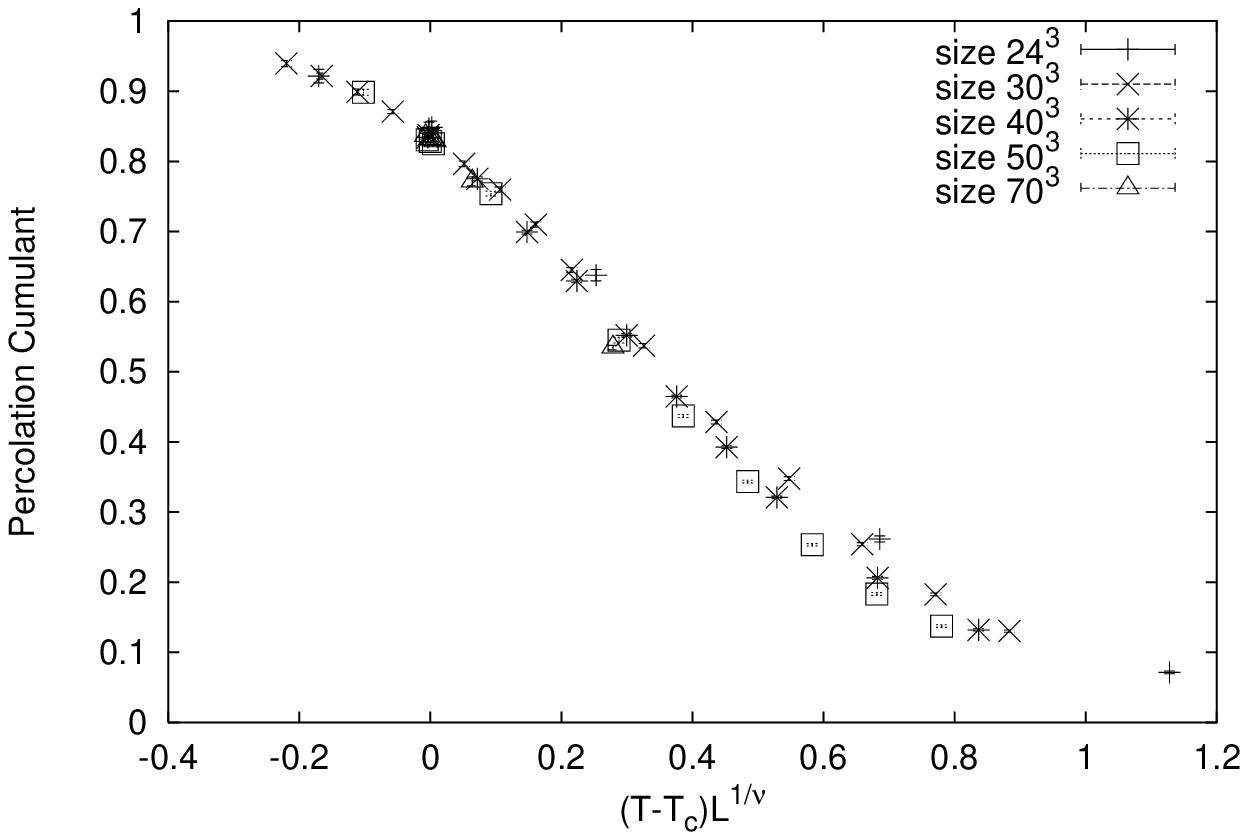,width=8.5cm}}
    \put(220,-10){\begin{minipage}[t]{7.5cm}{{\footnotesize
          Figure 6. Rescaled percolation cumulant for O(4)
          using $\beta_{c}=0.9359$ and the
          3-dimensional random percolation exponent $\nu=0.88$.}}
    \end{minipage}}
  \end{picture}

\newpage

\begin{center}
\begin{tabular}{|c||c|c|c|}
\hline
& $\beta_{c}$ & ${\beta}/{\nu}$ & ${\gamma}/{\nu}$ \\
\hline
Percolation results & 0.45418(2)& 0.516(5)
& 1.971(15) \\
\hline
Thermal results \cite{O2}& 0.454165(4)                  & 0.5189(3)
& 1.9619(5)\\
\hline
\end{tabular}\end{center}
\centerline{Table I. Comparison of the thermal and percolation thresholds and
  exponents for O(2) .}

\vskip1.2cm
\begin{center}
\begin{tabular}{|c||c|c|c|}
\hline
& $\beta_{c}$ & ${\beta}/{\nu}$ & ${\gamma}/{\nu}$ \\
\hline
Percolation results & 0.93595(3)& 0.515(5)
& 1.961(15) \\
\hline
Thermal results & 0.93590(5)\cite{Manf}                  & 0.5129(11)\cite{O4}
& 1.9746(38)\cite{O4}\\
\hline
\end{tabular}\end{center}
\centerline{Table II. Comparison of the thermal and percolation thresholds and
  exponents for O(4).}

\vskip1cm
The agreement with the physical values 
in Refs.\ \cite{O2,Manf,O4} is good.
So far we have presented the results obtained
using the first approach. The results derived using 
the second approach are essentially the same; besides,
we observe an improved quality of the scaling, mainly because 
of the use of periodic boundary conditions, which
reduce considerably the finite-size effects.
In particular we show
in Figs. 7, 8 
the scaling of $S$ and $P$ 
at the thermal thresholds reported in Refs.\ \cite{O2,Manf}.
We observe very small finite-size effects (lattices of $L\ge 20$ are
used in the fits), especially for the O(2) case, 
which is in contrast to what 
is observed for thermal observables \cite{engels}.
The slopes 
of the straight lines are in agreement with the 
values of the thermal exponents' ratios ${\beta}/{\nu}$,
${\gamma}/{\nu}$.

We thus confirm numerically the equivalence
found in Section 3 for the O(2) case, and verify that it holds also in
the O(4) case.
\vskip0.3cm
\setcounter{figure}{6}
\begin{figure}[h]
\begin{center}
\epsfig{file=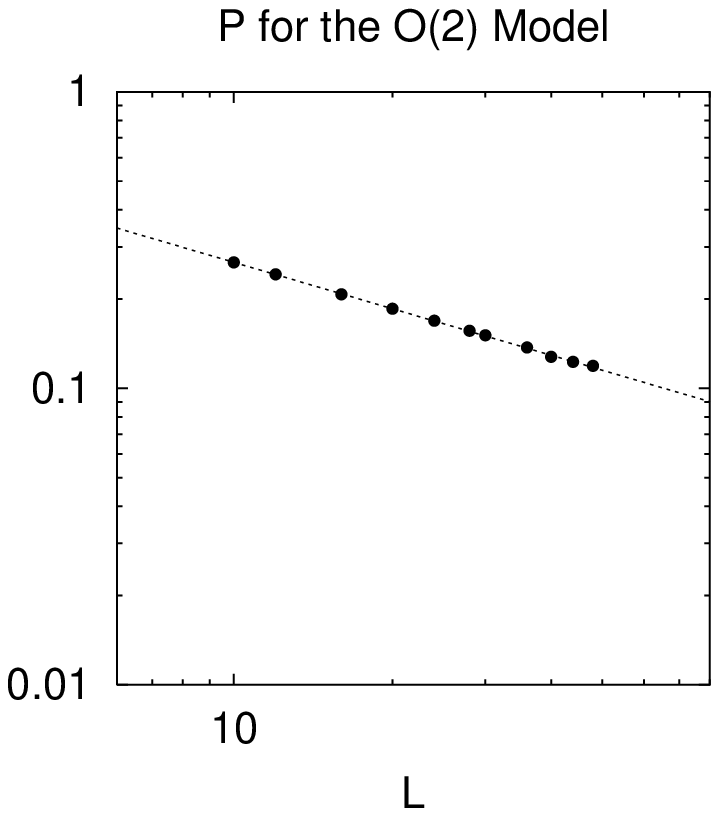,width=7.5cm}
\epsfig{file=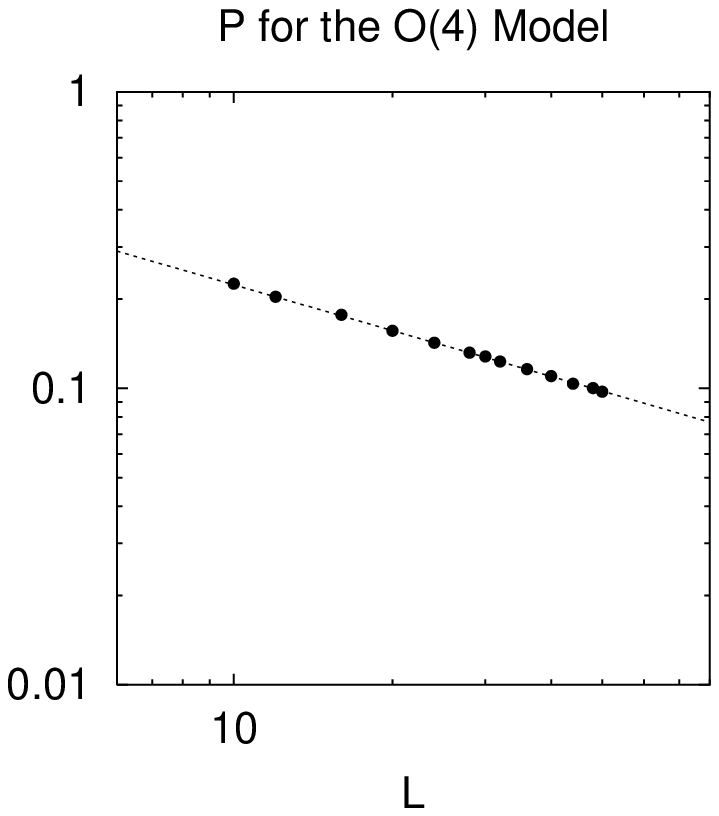,width=7.5cm}
\vskip0.3cm
\caption{
Finite-size-scaling plot at $T_c$ for the percolation observable $P$ as
a function of the lattice size $L$. The slopes in the plots correspond
to $\beta'/\nu' = 0.521(3), 0.513(6)$ respectively for O(2) and O(4).
Error bars are one standard deviation.
}
\label{figP}
\end{center}
\end{figure}
\begin{figure}[h]
\begin{center}
\epsfig{file=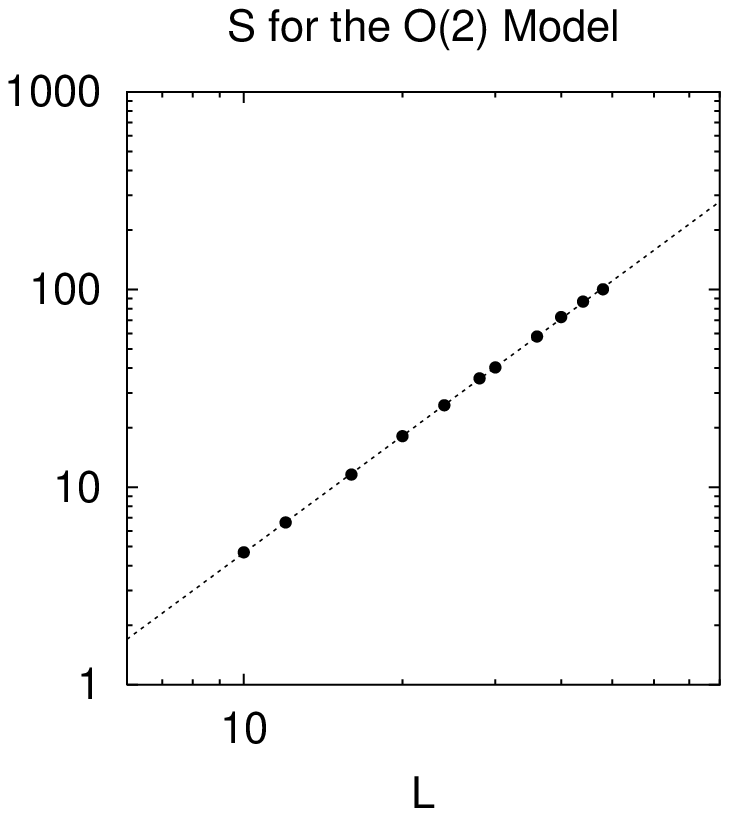,width=7.5cm}
\epsfig{file=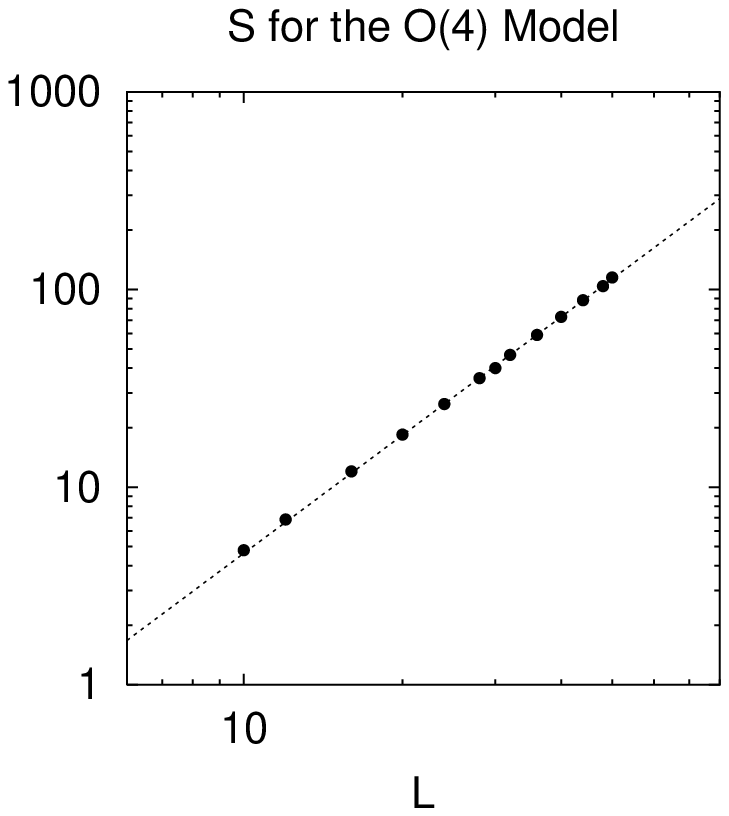,width=7.5cm}
\vskip0.3cm
\caption{
Finite-size-scaling plot at $T_c$ for the percolation observable $S$ as
a function of the lattice size $L$. The slopes in the plots correspond
to $\gamma'/\nu' = 1.97(1), 1.99(1)$ respectively for O(2) and O(4).
The two curves look surprisingly similar to each other.
Error bars are one standard deviation. 
}
\label{figS}
\end{center}
\end{figure}
\vskip-1cm
\section{Conclusions}

In this work we have shown that the spontaneous breaking of 
the continuous rotational symmetry for the 3-dimensional O(2), O(3)
and O(4) spin models can be described as percolation of 
Wolff clusters. For O(2) and O(3) the result was
proven analytically, for O(4)
it was derived by means of lattice Monte Carlo simulations. 
In all cases, the number $n$ of components of the spin vectors 
{\bf s} does not seem to play a role; the result is thus likely to be valid
for any O(n) model. 

\bigskip

\centerline{{\bf \large Acknowledgements}}

\bigskip

It is a pleasure to thank J. Engels for helpful discussions. 
We would also like to thank the TMR network ERBFMRX-CT-970122, 
the DFG Forschergruppe Ka 1198/4-1 and German Science 
Ministry BMBF under contract 06BI902 for financial support.

\end{document}